\documentclass[conference]{IEEEtran}
\usepackage[maxnames=20]{biblatex}
\usepackage{dirtytalk}
\usepackage{xcolor}
\usepackage{url}

\newcommand{\replicationPackage}{\url{https://doi.org/10.5281/zenodo.6273497}}
\newcommand{\catname}[1]{\textsf{\small #1}}

\usepackage[disable]{todonotes}
\usepackage{todonotes}
\presetkeys%
    {todonotes}%
    {inline}{}

\usepackage{booktabs}
\usepackage{multirow}
\usepackage{multicol}
\usepackage{makecell}

\usepackage[many]{tcolorbox}
\newtcolorbox{mybox}[1][]{
  breakable,
  title=#1,
  colback=white,
  colbacktitle=white,
  coltitle=black,
  fonttitle=\bfseries,
  bottomrule=0pt,
  toprule=0pt,
  leftrule=3pt,
  rightrule=3pt,
  titlerule=0pt,
  arc=0pt,
  outer arc=0pt,
  colframe=black,
}
\newcommand{\q}[1]{\emph{``#1''}}

\begin{document}

%
\title{A Case Study of Building Shared Understanding of Non-Functional Requirements in a Remote Software Organization}

 

\author{\IEEEauthorblockN{
Laura Okpara, 
Colin Werner,
Adam Murray,
Daniela Damian}
\IEEEauthorblockA{
\emph{Department of Computer Science}\\
\emph{University of Victoria, Victoria, Canada}\\
\emph{\{lauraokpara, colinwerner, adammurray, danielad\}@uvic.ca}}}


%


\maketitle

\begin{abstract}
Building a shared understanding of non-functional requirements (NFRs) is a known but understudied challenge in requirements engineering, especially in organizations that adopt continuous software engineering (CSE) practices. 
During the peak of the COVID-19 pandemic, many CSE organizations complied with working remotely due to the imposed health restrictions; some continued to work remotely while implementing business processes to facilitate team communication and productivity. In remote CSE organizations, managing NFRs becomes more challenging due to the limitations to team communication coupled with the incentive to deliver products quickly. While previous research has identified the factors that lead to a lack of shared understanding of NFRs in CSE, we still have a significant gap in understanding how CSE organizations, particularly in remote work, build a shared understanding of NFRs in their software development.
We conduct a three-month ethnography-informed case study of a remote CSE organization. Through thematic analysis of our qualitative data from interviews and observations, we identify a number of practices in developing a shared understanding of NFRs. The collaborative workspace the organization uses for remote interaction is Gather, which simulates physical workspaces, and which our findings suggest allows for informal communications instrumental for building shared understanding. As actionable insights, we discuss our findings in light of proactive practices that represent opportunities for software organizations to invest in building a shared understanding of NFRs in their development.

\hspace{2pt}

\end{abstract}

%
\IEEEpeerreviewmaketitle

\section{Introduction}
\label{sec: introduction}

Many software organizations are adopting continuous software engineering (CSE) practices, such as continuous integration and delivery \cite{johanssen2018practitioners}, to support the release of working software through automation and shorter cycle times between releases \cite{fitzgerald2017continuous}. CSE has implications for requirements engineering practices as it combines activities from agile methodologies that emphasize individuals and interactions, working software, customer collaboration, and response to rapid changes \cite{fowler2001agile}. 



However, previous research has identified the lack of shared understanding as a significant challenge in managing non-functional requirements (NFRs) in CSE \cite{werner2021continuously}. Shared understanding of NFRs is essential due to the complex, cross-cutting nature of NFRs \cite{ameller2012software} and the importance of NFRs to the success of software projects. 
For example, consider NFRs such as \textit{security} \cite{ullah2011survey} \cite{sen2015estimating}; e.g. if data encryption is poorly implemented, a data breach can expose user information. Werner et al. describe the factors that contribute to the lack of shared understanding of NFRs, such as the fast pace of change and lack of domain knowledge \cite{werner2020lack}. However, there is still a lack of empirical evidence regarding how a CSE organization builds and manages a shared understanding of NFRs. 

Working remotely creates further challenges in managing NFRs, along with their understanding, due to geographic distance, differences in context, and heavy reliance on communication technologies \cite{gibson2003virtual}. There are barriers to building a shared understanding, as conveying information through non-verbal communication is difficult in remote environments \cite{kniel2021riding}. During the COVID-19 pandemic, there was a global shift from in-person activities to working remotely \cite{bick2020work}. Many organizations had to adapt processes and systems to support team productivity, through virtual communication tools, social engagements, and peer support \cite{miller2021your}. 

With the significant shift of software organizations working remotely, the motivation for this study stems from the need to explore how remote CSE software organizations build a shared understanding of NFRs to potentially minimize rework \cite{werner2020lack} and other costs that may result from a lack of shared understanding of NFRs. 
The primary goal of this study is to investigate how remote CSE software organizations build and maintain a shared understanding of NFRs.
The secondary goal of this study is to provide additional insight into the limitations or challenges to a shared understanding of NFRs in CSE. Two main research questions guided our study:

\textbf{RQ1:} How does a \textit{remote} software organization that adopts CSE practices reach a shared understanding of NFRs?

\textbf{RQ2:} What are the limitations to the shared understanding of NFRs in a \textit{remote} software organization that adopts CSE practices?

Our work brings two significant contributions to the field of requirements engineering:
\begin{enumerate}
  \item We add to the growing knowledge of shared understanding of NFRs in requirements engineering by explicitly describing practices for building a shared understanding of NFRs in remote organizations that adopt CSE practices.
  \item We discuss practical ways for remote CSE organizations to proactively build a shared understanding of NFRs.
\end{enumerate}

\section{Background and Related Work}
\label{sec: related_work}
\subsection{Non-functional requirements}
Previous empirical studies suggest that managing NFRs can be challenging to software organizations for varying reasons, such as sparse documentation of NFRs or when documentation may be imprecise \cite{ameller2012software}, along with the lack of consensus on how to sufficiently document and communicate NFRs \cite{glinz2007non}. 
NFRs are important to the success of software projects; however, research suggests that there is still some conflict in how software organizations manage NFRs \cite{mairiza2010investigation}. NFRs are often neglected in requirements engineering due to the difficulties in eliciting NFRs, as NFRs have cross-cutting concerns with FRs \cite{ullah2011survey}. Research suggests that NFRs have a representation problem and when NFRs are not described or represented properly, they can be misinterpreted as FRs \cite{glinz2007non}. NFRs such as \textit{security} and \textit{privacy} are important to software projects, but often neglected \cite{ullah2011survey} and may be specified too late resulting in system vulnerabilities \cite{sen2015estimating}.
The main challenge is to understand the complex nature of NFRs and to create practices to manage the shared understanding of NFRs \cite{mairiza2009managing}. This is the gap that we intend to fill in this study by exploring practices to build a shared understanding of NFRs.
 
\subsection{Shared Understanding and CSE}
CSE increases the need for an effective flow between customer needs and the rapid delivery of a product or service \cite{fitzgerald2017continuous} through the automated release of working software in short release cycles. While this concept is already established through the agile manifesto \cite{fowler2001agile}, CSE is a more holistic approach as it includes activities to extend these practices such as continuous verification, improvement, and innovation \cite{fitzgerald2017continuous}. 

Previous research suggests that shared understanding among people exists in two forms: implicit and explicit \cite{glinz2015shared}. There is implicit shared understanding when people have a mutual understanding of unspecified facts or assumptions; whereas there is explicit shared understanding when people have a mutual understanding of specifications such as documents.
Shared understanding is beneficial to CSE projects as it can promote successful collaboration \cite{darch2009shared}; when shared understanding exists, a team member is able to predict other's behaviours, there is an increase in team motivation and less conflict or mistrust among the team. Continuous integration, as CSE emphasizes, provides quick feedback to support the frequent release of working software, ensuring that potentially failure-inducing problems are identified and resolved as quickly as possible \cite{fitzgerald2017continuous}. We bring further evidence from this study toward how shared understanding is built in CSE.

Werner et al. studied shared understanding of NFRs in CSE within three organizations \cite{werner2020lack}. They traced the effects of the lack of shared understanding in the form of rework in software projects, and they found that managing NFRs in CSE comes at a cost, such as fast pace of change, inadequate communication, and a lack of domain knowledge. They recommend practices for building a shared understanding of NFRs: communication and shared standards. Similarly, previous research reports the difficulties with automating NFRs and the problem of prioritizing functional requirements (FR) over NFRs \cite{werner2021continuously}. However, these studies do not address how a CSE organization builds a shared understanding of NFRs.

\subsection{Shared Understanding and Informal Communication}


Shared understanding cuts across several fields, disciplines, and contexts, such as education \cite{puntambekar2006analyzing}, medical science \cite{heasman2019neurodivergent}, design science \cite{cash2017supporting}, or engineering \cite{glinz2015shared}. Several definitions of shared understanding exist in literature, one study describes shared understanding as ``the ability of multiple agents to coordinate their behaviours with respect to each other to support the realization of common goals or objectives'' \cite{bittner2013shared}. Hinds conceptualizes shared understanding as a collective way of organizing relevant knowledge and a means for team members to anticipate and predict the behaviors of their peers or group \cite{hinds2003knowledge}. These definitions focus on people and how they communicate to reach a common goal or objective. 
Research evidence suggests that in software organizations, informal communication is essential for understanding and communicating about stakeholder values and needs \cite{beck2000extreme}. Informal communication is interactive and includes unplanned interactions that occur in the midst of daily activities \cite{torlind2002supporting} \cite{dorairaj2011effective}. Kraut et al. discuss informal communication as based on the degree of preplanning, therefore, informal communication can be scheduled, intended, opportunistic, or spontaneous \cite{kraut1990informal}.

Research evidence in requirements engineering suggests that inadequate communication contributes to a lack of shared understanding \cite{werner2020lack}. Furthermore, daily informal communication allows team members to develop working relationships and encourages a better flow of information \cite{agerfalk2005framework}. Having a virtual shared space in remote software organizations is important for collaboration on software projects \cite{balakrishnan2008visualizations}. In this study, we bring empirical evidence on \textit{how} a remote software organization manages communication of NFRs in the absence of traditional in-person informal communication.

\section{Methodology}
\label{sec: methodology}

Our case study used ethnography-informed methods to study how a remote software organization, Alpha (fictitious name to protect confidentiality), attempts shared understanding in their product development. We first outline the study setting in terms of the organization studied, its processes and working environment. We then follow with a description of the data collection and analysis methods.

\subsection{Study Setting}
We identified a remote organization, Alpha, that practices CSE. Alpha is a Canadian based software organization that develops loyalty, rewards, and referral programs for businesses, as their primary revenue source. Alpha has operated for 9 years with 29 global employees.
Alpha is primarily client-focused, with most of its products having heavy visual components, including third-party platforms and storing client information. Alpha prioritizes NFRs such as performance, usability, security, and maintainability to meet their stakeholders' needs. The software engineering team at Alpha includes project leads (manager or developer), solutions architects, front-end and back-end software developers, user experience (UI/UX) designers/developers, and software integration engineers.

During the COVID-19 pandemic, Alpha shifted to remote work to comply with the health restrictions imposed on businesses within the region and country and, having adapted its business operations and approach, stayed fully remote for over two years. Alpha used Gather, a video chat platform designed to make virtual interactions more human.


\emph{Alpha's communication methods and tools.}
Similar to many other organizations during the COVID-19 pandemic, Alpha had to pivot to work remotely. Alpha initially used video conferencing tools such as Zoom; however, it soon discovered that it did not create connectedness within the software team, e.g. \q{we started out like that when the pandemic first hit, we all went remote. Initially, it was so disconnected. No one felt connected to anyone else. It was so difficult to engage people in random conversation} (Manager).
Alpha then switched to using Gather \cite{gather_town}, a virtual interaction tool that creates a work environment similar to a physical workspace, including desk spaces, meeting rooms, and avatars that represent the team members. 
Gather is fully customizable and allows Alpha to interact virtually as much as they would do in a physical space: whenever an avatar (team member) "walk" near another team member's avatar, a video call would be initiated in an easy and ubiquitous way, to simulate informal communication available in physical co-location, e.g \q{Gather has made it possible for us as a remote team to have informal conversations. You literally just walk your little character up to someone else's little character, and you're talking to them.} (Manager) During meetings, Alpha's employees "walk" their avatar to the virtual meeting room in Gather and can choose to sit at the table. A group video call on Gather would begin as soon as they are near each other. Alpha's employees can also create private spaces (bubbles) with each other while in group meetings for one-on-one conversations when necessary, creating the feeling of being in the same space with each other. Alpha uses Gather to host social events, such as Christmas parties, encouraging team members to interact with each other beyond their daily work.  

Alpha also uses Slack \cite{slack} to provide and receive frequent updates relevant to daily activities, and web documentation tools to communicate and document requirements such as Coda \cite{coda_docs} to review backlogs, view the status of previous or ongoing tasks, and update tasks seamlessly throughout the life cycle of projects. 

\emph{Alpha's software and requirements engineering processes.} Alpha adopted CSE practices such as frequent feedback loops, continuous verification, and continuous testing based on stakeholder needs. Empirical evidence suggests an organization that adopts CSE may focus on certain CSE practices such as continuous integration for several reasons to suit their business and stakeholder needs \cite{elazhary2021uncovering}. The software engineering team at Alpha uses continuous delivery to frequently deliver working software in short cycles, welcoming changing requirements to meet the stakeholders' needs. In addition, Alpha implements a people-driven, autonomous approach incorporating tribes and guilds \cite{smite2019spotify}. A tribe describes a group of people who work on similar feature areas of the software products; they coordinate and align themselves through the tribe leader, usually a senior member of the tribe. Tribe members have the necessary software engineering skills to come together and engineer working software features from "end-to-end." A guild describes a group of people with similar interests who have similar roles or perform similar tasks. Guild members come together to share knowledge, support each other and share processes to create a community that fosters learning and growth in specific areas. 


Before starting a new project, a project kick-off meeting first occurs that includes stakeholders, such as success representatives, customers, or upper-level management. The meeting defines some of the high-level software requirements, including NFRs, and the project scope through user/data flows, product designs, and demos to understand the software product to be built.
The project evolves with the delivery of working usable software in short cycles by continuously verifying project requirements with stakeholders throughout the software development process \q{to ensure that the right thing has been built and it has been built right} (Developer). The project lead initiates frequent feedback cycles such as weekly or bi-weekly meetings with the stakeholders involved in the project, the feedback from these meetings can be change requests, new requirements or software bugs discovered. The rapid feedback cycles continue and the project continues to evolve until all requirement requests have been implemented or all of the stakeholders are satisfied with the product built. 

At Alpha, software developers have some freedom and flexibility to implement the software requirements, especially the NFRs. However, an engineer must collaborate with the project lead to remove ambiguity and to create an architecture allowing all components to work together. 


\subsection{Data Collection}
For three months, one of the authors was embedded at Alpha to learn about the products, processes, and business. Through many informal conversations with the software engineering team, an intention was to understand how the team shared understanding, knowledge and how they communicate and document requirements. Alpha onboarded the on-site researcher by providing a work email, providing a dedicated computer, and granting access (through invitation) to their collaboration tools and platforms such as Gather, Slack, Coda, Google Drive \cite{google_drive} and GitHub \cite{github}. Our data collection techniques were through observations and interviews.

\emph{Participant Observation:}
The on-site researcher observed the participants and their interactions during regular work activities and meetings with the software team and stakeholders, including customers and upper-level management. 
By engaging in informal conversations with employees, the researcher learned about their roles and experience and developed a working relationship with the team to help design the interview questions and interpretation of interview responses.

\begin{table}  
\centering  
 \caption{Interviewees}
 \label{tab:interview-participants} 
 \begin{tabular}{llll}
 \toprule      
 \textbf{Role} & \textbf{Work Area} & \textbf{Exp. in Org.} & \textbf{Overall Exp.} \\
 
 \midrule       
 Developer & Visual Dev & \textless  1 yr & \textless 1 yr \\
 Developer & Visual Dev & \textless 1 yr & \textless 1 yr \\
 Developer & Project Engr & \textless 1 yr & \textless 1 yr \\
 Developer & Visual Dev& \textless 1 yr & \textless 1 yr\\
 Developer & Infrastructure & \textless 1 yr & \textless 3 yr\\
 Developer & Infrastructure  & \textless 3 yr & \textless 3 yr\\
 Developer & Visual Dev & \textless 3 yr & \textless 3 yr\\
 Manager & Product & \textless 5 yr  & \textless 5 yr \\
 Manager & Project/Product & \textless 5 yr & \textless 10 yr\\
 Manager & Infrastructure & \textless 5 yr & \textless 20 yr\\
 Manager & Executive  & \textless 10 yr & \textless 10 yr \\

\bottomrule

\end{tabular}
\end{table}
\emph{Interviews:}
Eleven semi-structured interviews were conducted via Zoom during the third month of the study. Each interview time varied between 30 and 45 minutes and was fully transcribed.
The interviewees had various roles with varying years of experience, from a couple of months to fifteen years of experience. To protect the anonymity of our participants, we label developer/designer roles and team leads as developers and other remaining management roles as managers, as seen in Table~\ref{tab:interview-participants}. Following a grounded theory approach, we iteratively developed the interview questions based on emerging patterns that helped us to establish an area of focus or interest \cite{hoda2011grounded}. The interviews focused on highlighting how shared understanding of NFRs is established and what challenges may limit the shared understanding of NFRs. We used open-ended questions to allow participants to describe their experiences. We sought each participant's view on shared understanding and how Alpha may reach a shared understanding of NFRs. The interviews were audio-recorded with consent.
A sample of the interview questions can be found in Table \ref{tab:interview-questions}.

\begin{table}
    \centering
    \caption{Sample Interview Questions}
    \label{tab:interview-questions}
    \begin{tabular}{rp{7.5cm}}
    \toprule
    \textbf{Q\#} & \textbf{Question} \\ 
    \midrule 
    Q1 & How would you describe your role and the work you do at Alpha? \\
    Q2 & Do you recognize the difference between functional and non-functional requirements in software projects? \\
    Q3 & Based on your role and experience, which non-functional requirements are vital to the success of your projects at Alpha? \\
    Q4 & Have you heard of the term shared understanding? what does a mutual or shared understanding mean to you? \\
    Q5 & How would you describe the process of reaching a shared/mutual understanding of non-functional requirements for any project?\\
    Q6 & When can you say that everyone on your team understands the non-functional requirements sufficiently to produce substantial work? \\
    Q7 & How would you describe the factors that hinder or limit your understanding of non-functional requirements when working with other people? \\
\bottomrule
\end{tabular}
\end{table}

\subsection{Data Analysis}
Data analysis involved fully transcribing audio recordings from interview sessions and creating insights using the background information obtained through observations. The observation data collected was analyzed by noting patterns and data relevant to our study. We used this data to describe our study setting and design the research questions for the interviews. We also used the data from our observations to interpret the interview findings by understanding and describing Alpha's context and discussing our study's implications. 
We used the open and axial coding methods from grounded theory \cite{hoda2011grounded} to identify themes \cite{kiger2020thematic} across several codes from the interviews, while creating memos to describe the relationships and patterns observed in the data and any additional questions that may be explored. We also identified the triggers that led Alpha to build a shared understanding of NFRs, as suggested by Werner et al. \cite{werner2020lack}, further discussed in Section \ref{sec: discussion}.
Two independent coders were involved in the thematic coding process \cite{kiger2020thematic}.
Agreement sessions were used to consolidate coding strategies, and to establish the reliability of the codes.
We calculated the inter-rater reliability agreement until we reached substantial inter-rater reliability \textgreater 0.60, using the Cohen Kappa's coefficient for measuring observer agreement for categorical data \cite{landis1977measurement}.


\begin{table*}[ht]
    \centering
    \caption{Summary of Practices for Building a Shared Understanding of NFRs (RQ1)}
    \label{tab:codebook-examples}
    \begin{tabular}{llccc}
    \toprule
    \textbf{Practices}  & \textbf{Description / Key Phrases} 
    & \textbf{Implicit / Explicit} \cite{glinz2015shared}
    & \textbf{Triggers} \cite{werner2020lack}
    & \textbf{Proactive / Reactive}\\
    \midrule
    \multirow{3}{*}{\makecell[l]{\textbf{Validating NFRs} \\ \textbf{through feedback}}}
    & Validating NFRs through code review 
    & \multirow{3}{*}{Implicit}
    & \multirow{3}{*}{Rework}
    & \multirow{3}{*}{Proactive and Reactive} \\
    & Eliciting NFRs during feedback &&& \\
    & Capturing NFRs from demos &&& \\
    
    \midrule
    \multirow{3}{*}{\makecell[l]{\textbf{Deepening the understanding} \\ \textbf{of NFRs through experience}}}
    & Learning how the organization values NFRs
    & \multirow{3}{*}{Implicit/Explicit}
    & Change in scope & \multirow{3}{*}{Reactive} \\
    & Learning about NFRs over time && Rework & \\ 
    & Limitations due to lack of experience &&& \\
    
    \midrule
    \multirow{2}{*}{\makecell[l]{\textbf{Wireframing interface} \\ \textbf{designs}}}
    & Sketching solutions 
    & \multirow{3}{*}{Explicit}
    & \multirow{2}{*}{N/A} 
    & \multirow{2}{*}{Proactive} \\
    & Designing solutions using visual aids &&& \\ 
    
    \midrule
    \multirow{2}{*}{\makecell[l]{\textbf{Discussing problems} \\ \textbf{and solutions}}}
    & Meetings for implementing NFRs 
    & \multirow{3}{*}{Implicit}
    & Change in Scope & \multirow{2}{*}{Proactive and Reactive} \\
    & Informal conversations about problems && New requirements & \\
    
    \midrule
    \multirow{2}{*}{\textbf{Asking questions}} 
    & Asking questions about NFRs 
    & \multirow{3}{*}{Implicit}
    & Change in scope & \multirow{2}{*}{Proactive and Reactive} \\
    & Asking questions formally or informally && New requirements & \\
    
    \bottomrule
    \end{tabular}
\end{table*}

\section{Findings}
\label{sec: findings}

\subsection{RQ1: How does a remote software organization that adopts CSE practices reach a shared understanding of NFRs?}
\label{sec:rq1}
First, to understand how the software team views NFRs, we asked participants to tell us which NFRs were vital to the success of the projects that they work on. In our data analysis, we recognized three major NFRs that were mentioned repeatedly: usability, performance and maintainability. We also recognized a pattern where the NFRs mentioned during the interview sessions differed depending on the role and level of experience of the participant. Front-end developers and designers mainly describe usability and performance when referring to NFRs e.g. \q{Well, usually the biggest NFRs that we get are everything to do with how it looks. When you're dealing with visual specs, the most important thing is usually how it behaves in different environments} (Developer). Similarly, engineering leads and managers made reference to performance and understandability e.g. \q{But you could definitely consider there to be NFRs for our public APIs. And the things that you want to consider in those cases are, does the API make sense? Is it familiar for other developers? Is it something that they could easily discover?} (Manager). 
In another instance, a manager mentions security as an important NFR for their clients.
e.g. \q{They want to know that the data that they are providing us ... is secure, it's deleted when they need it to be deleted, and that the transmission is secure as well, that it's encrypted, and that there is no way to intercept the data as they're sending it to us or retrieving it} (Manager).

We then asked participants how they reach an understanding of NFRs for any project they have worked on. From our data analysis, we identified five practices (themes from our data) at Alpha, outlined in Table~ \ref{tab:codebook-examples}, and described below. In addition to the codes from which we developed these themes, Table~ \ref{tab:codebook-examples} also provides information as to whether our data indicates that the shared understanding was \emph{implicit} or \emph{explicit} \cite{glinz2015shared}, whether it was the result of \emph{triggers}, as found in the work of Werner and colleagues \cite{werner2020lack} and correspondingly whether they were \emph{proactive} rather than \emph{reactive}, i.e. no associated trigger: 
\begin{itemize}
 \item Validating NFRs through feedback
 \item Deepening the understanding of NFRs through experience
 \item Wireframing interface designs
 \item Discussing problems and solutions
 \item Asking questions
\end{itemize}

\subsubsection{Validating NFRs through feedback} At Alpha, feedback is provided and received during the software development process, as enabled by their use of standardized communication tools such as Slack and Gather. A manager describes a scenario where during a demo meeting, performance was captured as a valuable NFR. e.g. \q{I think we've actually seen this quite recently, with the demo project that's been happening here, where the engineering team was tasked with actually implementing the program. And by simply implementing it, you run up against all these NFRs that should exist. Because now you're going like, hey, I went to use it, it was really hard, I went to use it, it's taking forever to load} (Manager).

Alpha benefits from the quick feedback loops of CSE in the form of daily meetings, other scheduled meetings and informal conversations that happen during the software development process. These meetings mostly happen through Gather, the virtual office, e.g \q{We Slack or meet in Gather with screen-share and also every day, we have a morning meeting where we check-in and show what we're working on. And there can be opportunities for feedback there as well} (Developer). In another instance, a software developer describes the project kick-off meeting that helps with discovering NFRs like extensibility \q{Because when the 'shaping' is happening, we're starting to think about those NFRs about how it might work for other use cases} (Developer). Similarly, our participants express that there are processes that encourage the validation and capturing of NFRs such as code review and other feedback methods. e.g \q{Whereas the NFRs happen more passively. They're things that are built into processes, for example, when you create a pull request, there's a template that you have to fill out when you do a product launch ... And then obviously, there's things like code review, review and feedback from the managers that sort of catches the NFRs} (Developer). 


\subsubsection{Deepening the understanding of NFRs through experience} Four of our participants expressed to us that they deepen their understanding of NFRs over time through the projects that they work on within Alpha. These NFRs are sometimes not explicitly defined but incorporated into already existing processes and the developer learns how to implement them through those processes over a period of time. e.g \q{There's just a process that isn't formally presented, it's just picked up over time through training. I think that's the majority of NFRs are, for me anyways have been picked up that way} (Developer). Developers told us that a helpful way to deepen their understanding of NFRs is through learning how NFRs are handled in previous similar projects. e.g. \q{I think the biggest tool that has helped me learn NFRs has really just been reviewing previous work, reviewing previous specs, reviewing previous integrations, things like that, because it's the best way to get a sense of the expected code quality and expected way of doing things} (Developer). 

A developer describes how Alpha keeps people connected even while working remotely and how that helps with deepening the understanding of NFRs \q{I did mention earlier that Alpha is remote first but I do find that they do a good job of keeping people connected. And I think Gather town has really helped with that, it's just a place where you can feel like you're next to the people that you're working with. And they're, a few button clicks away from a call to talk about things and screenshare} (Developer). However, we observed that some of our participants expressed some difficulties in understanding NFRs due to a lack of experience. e.g \q{I still think I'm learning what the requirements are for spec writing, because it's just it's a nuanced aspect of the requirements .... So it's something that I'm still learning, but definitely have a better understanding of than when I first started.} (Developer). Nevertheless, a software developer describes learning Alpha's values as an important factor to gaining a shared understanding of NFRs e.g. \q{I guess the tricky thing is that a lot of getting shared understanding, I think is almost kind of unwritten, it's like, you have to get a sense for how your organization values those NFRs to be able to kind of work with others around you} (Developer).

\subsubsection{Wireframing interface designs}
We observed that another way Alpha reaches a shared understanding of NFRs was through wireframing. Our participants express that they mostly used Figma, sketches or minimal designs to describe some of the NFRs during project meetings or informal conversations. e.g. \q{And that usually involves using something like figma, or fig jam to quickly visualize what we're talking about. So really basic forms, like shapes, and just text, and blocks of lines that connect different screens} (Developer). To mitigate the challenges that could occur when working with visual components remotely, developers describe the use of screen-share on Gather, as a way that makes it easier to reach a shared understanding of NFRs. e.g \q{A kind of advantage of Gather is that you get to share your screen and [person] likes to write diagrams and stuff like that or just mock-up code on the screen, which definitely makes things easier to understand} (Developer).
A manager further describes the effectiveness of wireframing interface designs by stating \q{I think that's where things like 'shaping', wire-frames, starting to try to think through that user experience really helps. Design drawings, I find are really helpful. Early mock-ups help people because I find that a lot of the team does well with actually experiencing something. They quickly start to see ... this does the thing, but it's totally unusable for other reasons} (Manager). For NFRs such as usability, visualizing what needs to be built and receiving feedback on those designs from other developers and clients ensures that there is a shared understanding of those NFRs. e.g. \q{If we take the NFRs that are typically involved in something like visual design, in terms of user experience, obviously we have our designer who will design something, but we do try to obtain as much feedback on that design as possible. If possible, we actually like to communicate that or publicize it with the client, if there is something that's client-specific} (Manager).
 
\subsubsection{Discussing problems and solutions} 
Developers express that they sometimes gain more understanding about NFRs through discussions during regularly scheduled meetings on Gather, such as \textit{bi-weekly guild meetings}. e.g \q{ It's really just a time to talk about how we work on things, which is a great idea. And it's been helpful for talking about how we build things and non-functional requirements in specific areas of our jobs} (Developer). In another instance, a developer describes an experience of how he ensures that everyone who works on a project understands the NFRs when there is a change in scope. e.g \q{I'll set a meeting with everybody involved in the project. And if something changes, I'll make sure that everybody who is involved is together, and then we can actually hash it out and talk and then I'll usually take some kind of notes} (Developer).

In addition, participants also describe their communication through Gather, as being key to facilitating several feedback cycles and meetings where it feels like they are working face-to-face. e.g. \q{I can think of one software that we use, which is Gather. And Gather is really obviously a video communication tool. I think it does foster the ability for individuals to kind of interact more. And by interacting more, and by kind of stopping by somebody's desk that you may not have talked to in a while, you can kind of ensure that communication of different departments is kind of happening, and it's happening more organically} (Manager).
\begin{table*}[ht]
    \centering
    \caption{Limitations to Building a Shared Understanding of NFRs (RQ2)}
    \label{tab:limitations_NFRs}
    \begin{tabular}{lll}
    \toprule
    \textbf{Limitations}  & \textbf{Description / Key Phrases} & \textbf{Triggers} \\
    
    \midrule
    \multirow{3}{*}{\textbf{Gaps in Communication}} 
    & Making assumptions about a project/process & Redesign/Reshape of projects \\
    & Unclear communication of expectations & Scope creep \\
    && Change in scope \\
    
    \midrule
    \multirow{2}{*}{\textbf{Limited understanding of customer context}}
    & Learning about customers & \multirow{2}{*}{Scope Creep} \\
    & Insufficient understanding of customer needs & \\
    
    \midrule
    \multirow{4}{*}{\textbf{Individual differences}} 
    & Differences in ideas and background & \multirow{4}{*}{Rework} \\
    & Differences in skills and knowledge & \\
    & Differences in approach to work & \\
    
    \midrule
    \multirow{3}{*}{\textbf{Unspecified NFRs}} 
    & Uncommunicated NFRs & Difficulty implementing NFRs \\
    & Unwritten NFRs & New requirements \\
    & Unclear NFRs & \\
    
    \bottomrule
    \end{tabular}
\end{table*}

\subsubsection{Asking Questions}
Through several feedback methods, we observed participants using question asking to reach a shared understanding of NFRs. A manager describes asking questions to clients to elicit the NFRs that are important to them during the early stages of discussing the project. e.g. \q{you want to ask a lot of questions, you want to poke a lot of holes and things and, really go to the very edge of, you know, when you're sketching it out. So yeah, I would say that initially, it's a lot of questions} (Manager). In a separate instance, a developer uses question asking to clarify an NFR and ensure that all members of the team have a good understanding of what the goal is. e.g Concerning usability, \q{And you have to ask questions and make sure that the person fully understands what the goal is. Sometimes things can be unclear, but so long as you're constantly asking questions, then the person is confident that they know what they're doing} (Developer).

Furthermore, we asked participants about what methods of communication have been effective for reaching a shared understanding of NFRs. All of our participants mentioned Gather i.e, \q{And then we also have a virtual office environment called Gather, which is what we mainly use. I think Gather is good because it's more multimedia because you can talk and have video and share screens} (Developer). 

Finally, we asked participants when they know team members, inclusive of themselves, have a common or shared understanding of the NFRs. Developers describe this as when there is minimal feedback or questions during project meetings or conversations. e.g. \q{Like when we go into the 'deep-dive' we know when the non-functional requirements have been met, when we as a group meet, and there aren't gaps in our shared understanding. We come out of a review meeting and there's not a lot of feedback} (Developer). A manager describes this as when the team members are able to discuss the problems and solutions with each other and describe the behaviour of the software product e.g. \q{I'm kind of sufficiently satisfied with two conditions, what I think is a little bit more subjective, which is when the team members are starting to be able to explain the problem and understand the detailed nuance explain that problem to others. The other one is when you start to see like acceptance criteria that's really starting to look like one of our actual user acceptance tests. And by that point, you get a sense of if someone really understood the non-functional} (Manager). However, our participants acknowledge that shared understanding of NFRs should continue to evolve and there may not be a way to guarantee that there is a complete shared understanding of NFRs within the team at one time. A manager expresses this by saying \q{I don't actually think you can have [guarantee], honestly, until it's actually in a client's hands being used or users hands. Because, I mean, there's a chance we got the NFRs wrong, you know, every single person did} (Manager). In a different instance, a manager states that \q{I don't think it's possible to reach 100\% shared understanding, there's always going to be areas where people are running on assumptions, or people might have different interpretations} (Manager).

\subsection{RQ2: What are the limitations to a shared understanding of NFRs in a remote CSE software organization?} 
\label{sec:rq2}
We asked participants if they experienced any challenges with understanding NFRs when working collaboratively. From our data analysis, we found five main themes that describe the limitations to a shared understanding of NFRs: gaps in communication, limited understanding of customer context, individual differences and unspecified NFRs. Additionally, we identified the main triggers across these limitations. The themes, triggers, and associated codes are in Table~\ref{tab:limitations_NFRs}. 

\subsubsection{Gaps in communication}
In several instances, our participants describe gaps in communication within the software team and gaps in communication between the software team and the stakeholders due to unclear expectations for projects or assumptions about projects, which may lead to some rework or a change in the scope of a project. A manager describes a scenario in which the software team made inaccurate assumptions about what the stakeholders expected for a project regarding usability, which led to discussing additional requirements that were not within the scope of the project, \q{They had made the assumption that they would be able to go in and update these prints after launch. And we had made the interpretation that everything was just sort of hard-coded, write once and forget it. And so, in the end, it was something that we actually had to push back on the client and say that, making that something that's editable, would significantly increase the scope} (Manager). Another member of the software team describes a situation where there was some unclear communication regarding the project expectations for usability requirements which resulted in the developer working with inaccurate specifications. \q{There have been times where I've had a conversation with [person] about what a product needs to do and how he has envisioned it. And I guess some things just maybe got lost in translation. And I end up working on something that is different or works differently than what he had imagined. And vice versa} (Developer).

\subsubsection{Limited understanding of customer context}
A major challenge with reaching a shared understanding of NFRs when working with external customers is the limited understanding of the customer's business, products or needs. The lack of contextual knowledge about the customers can sometimes lead to misunderstandings or assumptions concerning the NFRs that are important to the customers, especially when an organization needs sufficient information about their customers or their customer's products to deliver a product successfully, as observed \q{if I'm unable to understand what the motivating use cases are if I'm unable to see how this project actually came to be, what the scope is even, I'm unable to go dive into what those NFRs are} (Developer). 
 
Similarly, a participant expressed having experienced difficulty when working on a project and not understanding extensibility as an NFR. This difficulty was mainly due to a limited understanding of the customer and their products. e.g.   \q{ I think a lot of the confusion probably also came from it being my first couple projects here. And so I was still wrapping my head around what our customers actually are, how we interact with them, the full range of our products, and how they're planning on implementing their programs} (Developer). 

\subsubsection{Individual differences} 
Differences in the background, skills, experience and ideas of team members can sometimes lead to misunderstandings of NFRs, especially when there is no common foundation of knowledge.
Alpha recognizes that having a diverse software team means that team members would have a different level of knowledge, skills and experience. e.g \q{the first thing that comes to mind when working with other team members, especially in a leadership position, is understanding that not every team member has the same level of technical or work experience as any other team member} (Manager). However, sometimes team members still make assumptions about how other members of the team think and how much they know and this may lead to a disparity in ideas about the product and how to implement the NFRs of the product. e.g. \q{But it just trickled down to things like the language we're using, just wasn't aligned even, people didn't even have the same ideas about what the functionality was. And that just like continued to trickle through the other non-functional requirements because it wasn't meeting some of the use cases anymore} (Developer).

\subsubsection{Unspecified NFRs}
We observed that some of our participants experienced difficulties with understanding or implementing NFRs due to unclear or unspoken expectations around these NFRs. Concerning maintainability, how a code-base can be understood and built upon with well-written documentation, a software developer describes difficulties. e.g \q{Definitely, for the specification writing for that project, it was pretty difficult for me, because I wasn't really sure what they should look like. And also there weren't great examples that I could base it off} (Developer). When NFRs are not clearly defined, communicated or documented, developers are often left with their assumptions on what NFRs to include for a project and how to implement the NFRs. Our participants expressed that this could happen, however, due to the frequent feedback loops within the software team, there are several opportunities to communicate and re-elicit the NFRs relevant to the work being done. e.g \q{The thing is non-functional requirements are not always communicated as much... trying to figure out what or not, understanding or communicating expectations around some of those NFRs from stakeholders. The good thing about developing at Alpha is that our loops are pretty quick, we have a lot of communication, so becomes quite clear when something isn't being met. And when that comes up, just work to try and elicit what those requirements are} (Developer).

\section{Discussion}
\label{sec: discussion}
CSE advocates for ongoing customer feedback and rapid iteration cycles \cite{fitzgerald2017continuous} \cite{glinz2015shared}, which should result in a higher level of understanding of customer requirements.
However, prior research indicates that CSE may pose a detrimental effect on the shared understanding of non-functional requirements \cite{werner2020lack}. 
In remote organizations, building a shared understanding becomes more challenging due to the additional recognized barriers with remote work, such as challenges with information sharing and team members having a reduced ability to build trustworthy relationships \cite{kniel2021riding}.

Our work explores the delicate balance of achieving a shared understanding of NFRs in a remote CSE organization (Alpha), and in this section, we discuss the findings of our empirical study.
Through our research, we uncovered five practices from Alpha to build a shared understanding of NFRs:  \catname{validating NFRs through feedback}, \catname{deepening the understanding of NFRs through experience}, \catname{wireframing interface designs}, \catname{discussing problems and solutions}, and \catname{asking questions}. 
We found some limitations within Alpha to building a shared understanding of NFRs, such as gaps in communication that may occur when team members make assumptions about each other's knowledge.
We discuss how Alpha mitigates some challenges with building a shared understanding in a remote setting through multiple collaboration tools, particularly Gather.

We believe these findings have important research implications concerning the intersection of shared understanding of NFRs and CSE in remote collaboration and how an organization could use practices that are proactive, rather than reactive, to build a shared understanding of NFRs. 

\subsection{Shared Understanding of NFRs in CSE}
Glinz describes implicit shared understanding as the mutual understanding of non-specified facts or assumptions \cite{glinz2015shared}. Our study suggests that Alpha relies on \emph{implicit} shared understanding of NFRs, partly due to CSE practices that encourage frequent feedback \cite{fitzgerald2017continuous} through formal, informal or face-to-face communication\cite{debbiche2014challenges} \cite{dorairaj2011effective}, as observed \q{a lot of getting to shared understanding is almost kind of unwritten} (Developer).

Our findings corroborate and bring additional, empirical detail about the amount of implicit shared understanding, as four of the five practices used to develop a shared understanding of NFRs were in fact implicit, namely \catname{validating NFRs through feedback}, \catname{deepening the understanding of NFRs through experience}, \catname{discussing problems and solutions}, and \catname{asking questions}. 
Only a single practice, \catname{wireframing interface designs}, was explicit in nature.

Werner et al. describe the importance of the shared understanding of configuration management as configurability is an NFR that also encompasses process quality \cite{werner2021continuously}, and supports the rapid development, configuration and deployment of software products. Prior research has described comprehensive configuration management \cite{humble2018continuous} for continuous integration and continuous delivery as a key component of CSE. Many organizations use more than a single configuration tool, such as Docker, to create a software environment and applications, as observed by Werner et al. \cite{werner2021continuously} within the three organizations in their study. An organization would benefit from continuous integration and delivery incorporated in CSE practices when they invest in building a shared understanding of NFRs such as configurability. Shared understanding is enhanced through the practices recommended by Werner et al. \cite{werner2020lack}, such as creating shared standards for configuration management and effective communication. At Alpha, developers use an in-house configuration tool for setting up their environment and application. Developers also use Process Street \cite{process_street}, Coda, and GitHub for documenting configuration standards and collaboration, as these tools allow the developers to coordinate their activities around configuration management. 

\subsection{Mitigating the effects of remote interaction on shared understanding in CSE}
Previous research analyzing surveys during the COVID-19 pandemic has suggested that effective communication is essential for a software team's productivity. Communication is often affected by changes in a team's culture and inefficient communication could pose a challenge to reaching milestones \cite{miller2021your}. As CSE organizations rely on frequent feedback cycles, the quality of communication within software teams may suffer when team members work from a distance, i.e. work remotely \cite{deshpande2016remote}.
However, our findings suggest that Alpha was able to mitigate some of the challenges in remote communication by using Gather, a virtual video conferencing tool that tries to simulate the face-to-face physical workspace \q{Gather has made it possible for us as a remote team to have informal conversations. You literally just walk your little character up to someone else's little character, and you're talking to them}. Gather, a collaboration tool, has not been heavily researched on collaborative work and, to the best of our knowledge, in collaboration in software development. Our findings bring evidence about Gather's value in creating spaces that approach in-person interaction that is conducive to shared understanding in software engineering. A 2020 study of a distributed design team working during COVID-19 lockdowns \cite{losev2020distributed} extensively used Zoom, Slack, and Google docs, identified difficulties with establishing a team spirit in a distributed setting due to the limited one-view per participant experience with Zoom meetings during group sketching and the difficulties with understanding the team member's progress. At Alpha, we believe that Gather has allowed Alpha to achieve the practices suggested by this study for mitigating the challenge of simulating a co-located design space and encouraging screen-sharing for remote collaboration. 

A key enabler for shared understanding is the ability to ask questions and receive feedback, regardless of background experience, education, or other demographic factors \cite{cash2017supporting}, that we identified as \catname{asking question} in our practices for building a shared understanding. Previous research suggests other enablers of shared understanding in remote teams such as visualizing information, online updates and trustworthiness \cite{kniel2021riding}. We further discuss our insights about these enablers at Alpha. First, we observed that Alpha uses wireframing for interface designs and screen-sharing to enable the team to set and understand the expectations for NFRs. This observation supports prior research on building shared understanding for supporting remote teams through building shared mental models with visual representations \cite{goldschmidt2007see}. 
Second, frequent communication ensures that there is a shared understanding of how team members feel and how the project progresses \cite{kniel2021riding}. Online updates like daily check-ins on Slack or formal or informal communication on Gather shows the importance of multiple forms of communication media \cite{jordan2016perceptions} \cite{hinds2003knowledge} and tools such as Gather, Slack, and file sharing. Third, mutual trust enables team members to collaborate and share feedback openly, thus improving shared understanding within remote teams \cite{kniel2021riding}. Previous research suggests that regular communication, including formal and informal communication, plays a vital role in creating relationships within remote teams and thus in building trust within the team, regardless of the difference in background, location or culture \cite{henttonen2005managing}. Alpha built trust within the team mainly through informal communication on Gather or Slack, where team members are open to seeking help from each other, as observed \q{If someone sends me a quick Slack message, and I look at it ... I will go straight to their desk and start to engage them in conversation so that we can whiteboard it or work through the problem}.

\subsection{How can organizations be proactive in building a shared understanding of NFRs?}
Through observations and the practices we identified in our study, we bring empirical evidence that further adds to and enhances the work by Werner et al.\cite{werner2020lack} on the \emph{recommended} practices for building shared understanding i.e., how team communication and shared development standards form the fundamental base for building a shared understanding.
In addition, we have identified a new practice of building a shared understanding through experience.
Our first practice, \catname{discussing problems and solutions}, is part of utilizing communication to develop a shared understanding.
Developers can also initiate the creation of a shared understanding by \catname{asking questions}, which is another form of communication.
\catname{Validating NFRs through feedback} is a form of communication that can happen through the continuous pipeline or in a meeting; regardless of \textit{how} that communication takes place, the emphasis is on the continuous validation of NFRs that can build a shared understanding.
An organization can leverage wireframes for interface designs, a form of shared development standards, that may enhance the proliferation of a shared understanding.
Finally, developers can deepen their understanding of NFRs through experience at their organization and how that particular organization values NFRs.

Werner et al. \cite{werner2020lack} observed that an organization builds a shared understanding of NFRs primarily due to a reactive response to accidental lack of shared understanding, related to various reasons -- \emph{triggers} -- such as scope creep, rework, regulatory requirements, accumulating technical debt, needs of an important customer, or a disruption of service.
In Table~\ref{tab:codebook-examples}, we show the triggers that emerged from our data analysis and how they correspond to the identified practices. In the practices we observed at Alpha, the \emph{deepening the understanding of NFRs through experience} is an example of such a reactive response, or where our data does not indicate that the response was the result of any of these or other triggers. 

However, we categorized four of the five practices we identified at Alpha as being \emph{proactive} in many instances, even though they were \emph{reactive} in a few instances. 
One practice in particular sheds more light on how Alpha adapted to working and building shared understanding in remote settings: \emph{wireframing interface designs}. 
Prior research suggests that using virtual shared spaces that enable the team to visualize solutions and interact with visualizations encourages better communication and enables shared understanding \cite{balakrishnan2008visualizations}. Through Gather's virtual shared space and screen-sharing feature, Alpha used tools such as Figma, an online whiteboard for visual representations with elements like blocks and shapes, to support collaboration with team members, thereby enabling the team to develop a shared understanding more quickly, as observed in prior research  \cite{balakrishnan2008visualizations}. 


We observed that a proactive effort to build a shared understanding may not require a formal process or documentation.
At Alpha, \q{There's a [guild\_name] guild and we meet once every two weeks to sort of check-in on NFRs. It's what practices are we going to try and use? What's working, what's not working? It's been helpful for talking about how we build things and NFRs in specific areas of our jobs.} The loose agenda of these meetings is rather informal and lends to the creative discussions around NFRs.
Our findings indicate that tools such as Gather offer a medium for these informal, yet important, conversations to occur. Developers at Alpha unanimously describe Gather as a method that has created a sense of being connected to each other, despite their remote setting. \q{I think Gather town has really helped with that, it's just a place where you can feel like you're next to the people that you're working with}.


Our findings have practical implications to software organizations as they show that while there is still a considerable amount of shared understanding built through reactive measures, there is also a sizable amount that is a direct result of some proactive practices.
One developer indicated that he found some NFRs as follows: \q{as you're going down into more details, during something like a journey map, that's where you start bumping into the non-functional requirements that may not have been clearly identified}.
However, when it comes to such NFRs, discussions did not happen until much later, as one participant noted, \q{that's when you really start the back and forth communication on how you're going to address them...So I think it's during that process of drilling down into the details on how a project is going to be run, that you start hitting them, and the NFRs}. Despite the complexity and importance of NFRs \cite{ullah2011survey} \cite{ameller2012software}, previous research observes that software developers discover NFRs during or after the software product implementation \cite{matoussi2008survey}. Eliciting NFRs at this stage may pose a challenge that could affect the entire software development process or lead to delays in software \cite{matoussi2008survey} \cite{ullah2011survey}; therefore, it is prudent for an organization to ensure the conversations about NFRs are occurring early enough to facilitate building a shared understanding. The definition of \emph{when} is left for future work; however, given the effects of architectural decisions on NFRs (or lack thereof), a shared understanding should be continually built throughout the development cycle, perhaps allotting enough time for ``just-in-time'' engineering \cite{ernst_case_2012}.

\section{Threats to Validity}
\label{sec: threats_to_validity}

We acknowledge the limitations of our qualitative study through four components of the total quality framework \cite{roller2015applied}: credibility, analyzability, transparency and usefulness. 

For the \textit{credibility}, the selection of Alpha may suffer from sampling bias as we focused on an organization willing to partner with us; however, we verified that their practices align with CSE through observation by one of our researchers. 
In addition, we interviewed participants with various roles and levels of experience.
To limit conscious or unconscious bias during data collection, we explained that study participation was completely anonymous and would not cause any consequence or risk to the participants. 
We started each interview session by exploring the definition of shared understanding. 
Afterwards, we explained shared understanding in the context of our research with examples to ensure that each participant had the same level of knowledge about shared understanding. 

For \textit{analyzability}, we used a transcription tool on the recorded audio from the interviews and one of the authors listened and verified each transcript. 
We used thematic analysis through the open, axial, and selective coding process from grounded theory \cite{hoda2011grounded}; we describe this process in Section~\ref{sec: methodology}. 
We used the inter-rater agreement process \cite{landis1977measurement} to align our codes and categories within our coding scheme. 
We used reflective memos during the coding process to describe patterns observed in the data collected for reference.

For \textit{transparency}, we provide a replication package (\replicationPackage) with our codebook, interview questions, and data about our inter-rater agreement sessions; however confidential data are excluded due to our non-disclosure agreement. 
Although we believe we reached saturation of codes in our data analysis, one limitation to our coding process may be the subjectivity of the exhaustiveness of our codes which may influence research results. 

For \textit{usefulness}, we do not propose that our results and findings hold true for all remote CSE organizations. 
We acknowledge that our single case study research provides results limited to our partner organization. 
However, to increase the usefulness of our findings, more case studies like ours would be helpful with a focus on how remote software organizations can proactively build and maintain a shared understanding of NFRs, and how to evaluate a shared understanding of NFRs.

\section{Conclusions and Future Work}
A shared understanding of NFRs is important for the success of many software projects. However, managing NFRs can pose a significant challenge to software organizations due to factors such as the conflicting nature of NFRs and the lack of shared understanding of NFRs. In addition, shared understanding of NFRs in CSE is still understudied and there is insufficient literature to show how organizations can effectively build a shared understanding. 

While prior research has explored how NFRs are managed in CSE, considering the challenges and best practices for managing NFRs, 
there is still the question of \textit{how} CSE organizations can build a shared understanding of NFRs. This paper describes the practices for building a shared understanding of NFRs, through a case study of a remote CSE organization. Through discussing the practices for building a shared understanding of NFRs in CSE, we observe that CSE organizations are proactively building a shared understanding, although still reactively building a shared understanding in several instances. Furthermore, we discuss the limitations to a shared understanding of NFRs in CSE, highlighting the similarities and differences between the limitations described in our study and prior research. 

With respect to the practical implications of our study, we acknowledge the limitations to our research as a single-case study, however, we believe that our research adds valuable empirical evidence to this research area and is a useful starting point to further explore how CSE organizations can build a shared understanding of NFRs. In future research, we seek to evaluate the effectiveness of proactive practices for building a shared understanding of NFRs in CSE. 


\section*{Acknowledgments}
We thank our partner organization and employees for their time and collaboration. We acknowledge Nowshin Nawar Arony and Neha Koulecar (University of Victoria) for their assistance in our study. Our research was supported by the Natural Sciences and Engineering Research Council of Canada (NSERC).



%



\AtNextBibliography{\small}
\printbibliography

@article{glinz2015shared,
  title={On shared understanding in software engineering: an essay},
  author={Glinz, Martin and Fricker, Samuel A},
  journal={Computer Science-Research and Development},
  volume={30},
  number={3},
  pages={363--376},
  year={2015},
  publisher={Springer}
}

@inproceedings{ernst_case_2012,
	Author = {Ernst, Neil A. and Murphy, Gail C.},
	Booktitle = {2012 {Second} {IEEE} {International} {Workshop} on {Empirical} {Requirements} {Engineering} ({EmpiRE})},
	Doi = {10.1109/EmpiRE.2012.6347678},
	Month = sep,
	Note = {ISSN: 2329-6356},
	Pages = {25--32},
	Title = {Case studies in just-in-time requirements analysis},
	Year = {2012}}

@article{fitzgerald2017continuous,
  title={Continuous software engineering: A roadmap and agenda},
  author={Fitzgerald, Brian and Stol, Klaas-Jan},
  journal={Journal of Systems and Software},
  volume={123},
  pages={176--189},
  year={2017},
  publisher={Elsevier}
}

@inproceedings{bittner2013shared,
  title={Why shared understanding matters--Engineering a collaboration process for shared understanding to improve collaboration effectiveness in heterogeneous teams},
  author={Bittner, Eva Alice Christiane and Leimeister, Jan Marco},
  booktitle={2013 46th Hawaii International Conference on System Sciences},
  pages={106--114},
  year={2013},
  organization={IEEE}
}

@article{cash2017supporting,
  title={Supporting the development of shared understanding in distributed design teams},
  author={Cash, Philip and Dekoninck, Elies A and Ahmed-Kristensen, Saeema},
  journal={Journal of engineering design},
  volume={28},
  number={3},
  pages={147--170},
  year={2017},
  publisher={Taylor \& Francis}
}

@article{smite2019spotify,
  title={Spotify guilds: how to succeed with knowledge sharing in large-scale agile organizations},
  author={Smite, Darja and Moe, Nils Brede and Levinta, Georgiana and Floryan, Marcin},
  journal={IEEE Software},
  volume={36},
  number={2},
  pages={51--57},
  year={2019},
  publisher={IEEE}
}

@article{fowler2001agile,
  title={The agile manifesto},
  author={Fowler, Martin and Highsmith, Jim and others},
  journal={Software development},
  volume={9},
  number={8},
  pages={28--35},
  year={2001},
  publisher={[San Francisco, CA: Miller Freeman, Inc., 1993-}
}

@article{elazhary2021uncovering,
  title={Uncovering the benefits and challenges of continuous integration practices},
  Doi={10.1109/TSE.2021.3064953},
  author={Elazhary, Omar and Werner, Colin and Li, Ze Shi and Lowlind, Derek and Ernst, Neil A and Storey, Margaret-Anne},
  journal={IEEE Transactions on Software Engineering},
  year={2021},
  publisher={IEEE}
}

@article{werner2021continuously,
  title={Continuously managing NFRs: Opportunities and challenges in practice},
  author={Werner, Colin and Li, Ze Shi and Lowlind, Derek and Elazhary, Omar and Ernst, Neil A and Damian, Daniela},
  journal={IEEE Transactions on Software Engineering},
  year={2021},
  publisher={IEEE}
}

@inproceedings{werner2020lack,
  title={The Lack of Shared Understanding of Non-Functional Requirements in Continuous Software Engineering: Accidental or Essential?},
  Doi={10.1109/RE48521.2020.00021},
  author={Werner, Colin and Li, Ze Shi and Ernst, Neil and Damian, Daniela},
  booktitle={2020 IEEE 28th International Requirements Engineering Conference (RE)},
  pages={90--101},
  year={2020},
  organization={IEEE}
}

@article{hinds2003knowledge,
  title={Knowledge sharing and shared understanding in virtual teams},
  author={Hinds, Pamela J and Weisband, Suzanne P},
  journal={Virtual teams that work: Creating conditions for virtual team effectiveness},
  pages={21--36},
  year={2003}
}

@article{kiger2020thematic,
  title={Thematic analysis of qualitative data: AMEE Guide No. 131},
  author={Kiger, Michelle E and Varpio, Lara},
  journal={Medical teacher},
  volume={42},
  number={8},
  pages={846--854},
  year={2020},
  publisher={Taylor \& Francis}
}

@inproceedings{darch2009shared,
  title={Shared understanding of end-users' requirements in e-Science projects},
  author={Darch, Peter and Carusi, Annamaria and Jirotka, Marina},
  booktitle={2009 5th IEEE International Conference on E-Science Workshops},
  pages={125--128},
  year={2009},
  organization={IEEE}
}

@article{puntambekar2006analyzing,
  title={Analyzing collaborative interactions: Divergence, shared understanding and construction of knowledge},
  author={Puntambekar, Sadhana},
  journal={Computers \& education},
  volume={47},
  number={3},
  pages={332--351},
  year={2006},
  publisher={Elsevier}
}

@article{heasman2019neurodivergent,
  title={Neurodivergent intersubjectivity: Distinctive features of how autistic people create shared understanding},
  author={Heasman, Brett and Gillespie, Alex},
  journal={Autism},
  volume={23},
  number={4},
  pages={910--921},
  year={2019},
  publisher={SAGE Publications Sage UK: London, England}
}

@inproceedings{hoda2011grounded,
  title={Grounded theory for geeks},
  author={Hoda, Rashina and Noble, James and Marshall, Stuart},
  booktitle={Proceedings of the 18th Conference on Pattern Languages of Programs},
  pages={1--17},
  year={2011}
}

@inproceedings{johanssen2018practitioners,
  title={Practitioners' eye on continuous software engineering: An interview study},
  author={Johanssen, Jan Ole and Kleebaum, Anja and Paech, Barbara and Bruegge, Bernd},
  booktitle={Proceedings of the 2018 International Conference on Software and System Process},
  pages={41--50},
  year={2018}
}

@inproceedings{glinz2007non,
  title={On non-functional requirements},
  author={Glinz, Martin},
  booktitle={15th IEEE International Requirements Engineering Conference (RE 2007)},
  pages={21--26},
  year={2007},
  organization={IEEE}
}

@article{landis1977measurement,
  title={The measurement of observer agreement for categorical data},
  author={Landis, J Richard and Koch, Gary G},
  journal={Biometrics},
  pages={159--174},
  year={1977},
  publisher={JSTOR}
}

@book{roller2015applied,
  title={Applied qualitative research design: A total quality framework approach},
  author={Roller, Margaret R and Lavrakas, Paul J},
  year={2015},
  publisher={Guilford Publications}
}

@inproceedings{ameller2012software,
  title={How do software architects consider non-functional requirements: An exploratory study},
  author={Ameller, David and Ayala, Claudia and Cabot, Jordi and Franch, Xavier},
  booktitle={2012 20th IEEE International Requirements Engineering Conference (RE)},
  pages={41--50},
  year={2012},
  organization={IEEE}
}

@inproceedings{mairiza2010investigation,
  title={An investigation into the notion of non-functional requirements},
  author={Mairiza, Dewi and Zowghi, Didar and Nurmuliani, Nurie},
  booktitle={Proceedings of the 2010 ACM Symposium on Applied Computing},
  pages={311--317},
  year={2010}
}

@inproceedings{mairiza2009managing,
  title={Managing conflicts among non-functional requirements},
  author={Mairiza, Dewi and Zowghi, Didar and Nurmuliani, Nurie},
  booktitle={Australian Workshop on Requirements Engineering},
  year={2009},
  organization={University of Technology, Sydney}
}

@book{gibson2003virtual,
  title={Virtual teams that work: Creating conditions for virtual team effectiveness},
  author={Gibson, Cristina B and Cohen, Susan G},
  year={2003},
  publisher={John Wiley \& Sons}
}

@inproceedings{miller2021your,
  title={“How Was Your Weekend?” Software Development Teams Working From Home During COVID-19},
  author={Miller, Courtney and Rodeghero, Paige and Storey, Margaret-Anne and Ford, Denae and Zimmermann, Thomas},
  booktitle={2021 IEEE/ACM 43rd International Conference on Software Engineering (ICSE)},
  pages={624--636},
  year={2021},
  organization={IEEE}
}

@misc{bick2020work,
  title={Work from home after the COVID-19 Outbreak},
  author={Bick, Alexander and Blandin, Adam and Mertens, Karel and others},
  journal={Working Papers 2017},
  year={2020},
  publisher={Federal Reserve Bank of Dallas, Research Department}
}

@article{deshpande2016remote,
  title={Remote working and collaboration in agile teams},
  author={Deshpande, Advait and Sharp, Helen and Barroca, Leonor and Gregory, Peggy},
  journal={International Conference on Information Systems, ICIS},
  year={2016},
  publisher={AIS Electronical Library (2016)}
}

@article{agerfalk2005framework,
  title={A framework for considering opportunities and threats in distributed software development},
  author={Agerfalk, Par J and Fitzgerald, Brian and Holmstrom Olsson, Helena and Lings, Brian and Lundell, Bjorn and {\'O} Conch{\'u}ir, Eoin},
  journal={Proceedings of the International Workshop on Distributed Software Development, Paris},
  volume={29},
  year={2005},
  pages ={47-61},
  publisher={Austrian Computer Society}
}

@inproceedings{torlind2002supporting,
  title={Supporting Informal Communication in Distributed Engineering Design Teams},
  author={T{\"o}rlind, Peter and Larsson, Andreas},
  booktitle={International CIRP Design Seminar: 16/05/2002-18/05/2002},
  year={2002}
}

@inproceedings{dorairaj2011effective,
  title={Effective communication in distributed Agile software development teams},
  author={Dorairaj, Siva and Noble, James and Malik, Petra},
  booktitle={International Conference on Agile Software Development},
  pages={102--116},
  year={2011},
  organization={Springer}
}

@book{beck2000extreme,
  title={Extreme programming explained: embrace change},
  author={Beck, Kent},
  year={2000},
  publisher={Addison-Wesley Professional}
}

@inproceedings{kraut1990informal,
  title={Informal communication in organizations: Form, function, and technology},
  author={Kraut, Robert E and Fish, Robert S and Root, Robert W and Chalfonte, Barbara L},
  booktitle={Human reactions to technology: Claremont symposium on applied social psychology},
  pages={145--199},
  year={1990}
}

@article{henttonen2005managing,
  title={Managing distance in a global virtual team: the evolution of trust through technology-mediated relational communication},
  author={Henttonen, Kaisa and Blomqvist, Kirsimarja},
  journal={Strategic Change},
  volume={14},
  number={2},
  pages={107--119},
  year={2005},
  publisher={Wiley Online Library}
}

@article{sen2015estimating,
  title={Estimating the contextual risk of data breach: An empirical approach},
  author={Sen, Ravi and Borle, Sharad},
  journal={Journal of Management Information Systems},
  volume={32},
  number={2},
  pages={314--341},
  year={2015},
  publisher={Taylor \& Francis}
}

@inproceedings{ullah2011survey,
  title={A survey on issues in non-functional requirements elicitation},
  author={Ullah, Saeed and Iqbal, Muzaffar and Khan, Aamir Mehmood},
  booktitle={International Conference on Computer Networks and Information Technology},
  pages={333--340},
  year={2011},
  organization={IEEE}
}

@article{kniel2021riding,
  title={Riding the Same Wavelength: Designers’ Perceptions of Shared Understanding in Remote Teams},
  author={Kniel, Jonas and Comi, Alice},
  journal={SAGE Open},
  volume={11},
  number={3},
  pages={21582440211040129},
  year={2021},
  publisher={SAGE Publications Sage CA: Los Angeles, CA}
}

@article{jordan2016perceptions,
  title={Perceptions of success in virtual cross-disciplinary design teams in large multinational corporations},
  author={Jordan, Shawn and Adams, Robin},
  journal={CoDesign},
  volume={12},
  number={3},
  pages={185--203},
  year={2016},
  publisher={Taylor \& Francis}
}

@article{goldschmidt2007see,
  title={To see eye to eye: the role of visual representations in building shared mental models in design teams},
  author={Goldschmidt, Gabriela},
  journal={CoDesign},
  volume={3},
  number={1},
  pages={43--50},
  year={2007},
  publisher={Taylor \& Francis}
}

@inproceedings{balakrishnan2008visualizations,
  title={Do visualizations improve synchronous remote collaboration?},
  author={Balakrishnan, Aruna D and Fussell, Susan R and Kiesler, Sara},
  booktitle={Proceedings of the SIGCHI Conference on Human Factors in Computing Systems},
  pages={1227--1236},
  year={2008}
}

@inproceedings{losev2020distributed,
  title={Distributed synchronous visualization design: Challenges and strategies},
  author={Losev, Tatiana and Storteboom, Sarah and Carpendale, Sheelagh and Knudsen, S{\o}ren},
  booktitle={2020 IEEE Workshop on Evaluation and Beyond-Methodological Approaches to Visualization (BELIV)},
  pages={1--10},
  year={2020},
  organization={IEEE}
}

@inproceedings{matoussi2008survey,
  title={A survey of non-functional requirements in software development process},
  author={Matoussi, Abderrahman and Laleau, R{\'e}gine},
  booktitle={Technical report TR-LACL-2008-7, University of Paris-Est (Paris 12)},
  year={2008},
  organization={LACL (Laboratory of Algorithms, Complexity and Logic)}
}

@misc{gather_town,
	title = {Gather - A better way to meet online},
	url = {https://www.gather.town/},
% 	urldate = {2020-05-22},
}

@misc{slack,
	title = {Slack is where the future works},
	url = {https://slack.com/},
% 	urldate = {2020-05-22},
}

@misc{coda_docs,
	title = {Coda | The doc that brings it all together.},
	url = {https://coda.io/},
% 	urldate = {2020-05-22},
}

@misc{github,
	title = {GitHub: Where the world builds software . GitHub},
	url = {https://github.com/},
% 	urldate = {2020-05-22},
}

@misc{google_drive,
	title = {Cloud storage for work and home - Google Drive},
	url = {https://www.google.com/intl/en_ca/drive/},
% 	urldate = {2020-05-22},
}

@misc{process_street,
	title = {Process Street | Checklist, Workflow and SOP Software},
	url = {https://www.process.st/},
% 	urldate = {2020-05-22},
}

@inproceedings{debbiche2014challenges,
  title={Challenges when adopting continuous integration: A case study},
  author={Debbiche, Adam and Dien{\'e}r, Mikael and Berntsson Svensson, Richard},
  booktitle={International Conference on Product-Focused Software Process Improvement},
  pages={17--32},
  year={2014},
  organization={Springer}
}

@article{humble2018continuous,
  title={Continuous delivery sounds great, but will it work here?},
  author={Humble, Jez},
  journal={Communications of the ACM},
  volume={61},
  number={4},
  pages={34--39},
  year={2018},
  publisher={ACM New York, NY, USA}
}

\end{document}